\documentclass[pra,amsfonts,superscriptaddress,showpacs,amsmath,twocolumn]{revtex4}
\usepackage{epsfig}
\usepackage{bbm}
\usepackage{multirow}
\usepackage{color}
\usepackage{hyperref}
\usepackage{pxfonts}
\usepackage{braket}

\begin{document}

\title{Counting Statistics of Many-Particle Quantum Walks} 

\author{Klaus Mayer}
\affiliation{Physikalisches Institut, Albert--Ludwigs--Universit\"at %
Freiburg, Hermann--Herder--Strasse~3, D--79104 Freiburg, Germany}

\author{Malte C. Tichy}
\affiliation{Physikalisches Institut, Albert--Ludwigs--Universit\"at %
Freiburg, Hermann--Herder--Strasse~3, D--79104 Freiburg, Germany}

\author{Florian Mintert}
\affiliation{Physikalisches Institut, Albert--Ludwigs--Universit\"at %
Freiburg, Hermann--Herder--Strasse~3, D--79104 Freiburg, Germany}
\affiliation{Freiburg Institute for Advanced Studies, Albert-Ludwigs-Universit\"at %
Freiburg, Albertstrasse 19, D-79104 Freiburg, Germany}

\author{Thomas Konrad}
\affiliation{Quantum Research Group, School of Pure and Applied Physics, University of KwaZulu-Natal, Private Bag 54001, Durban 4000, South Africa}

\author{Andreas Buchleitner}
\affiliation{Physikalisches Institut, Albert--Ludwigs--Universit\"at %
Freiburg, Hermann--Herder--Strasse~3, D--79104 Freiburg, Germany}

\date{\today}
\begin{abstract}
We study quantum walks of many non-interacting particles on a beam splitter array, as a paradigmatic testing ground for the competition of single- and many-particle interference in a multi-mode system. We derive a general expression for multi-mode particle-number correlation functions, valid for bosons and fermions, and infer pronounced signatures of many-particle interferences in the counting statistics.
\end{abstract}

\maketitle
\section{Introduction}
Quantum walks (QWs) provide the natural - yet unitary - extension of the classical random walk (RW) to the quantum regime \cite{PhysRevA.48.1687, Kempe:2003fk}, where the superposition principle allows a wave function to explore different paths simultaneously. The resulting interference in the transition amplitudes is cause to effects like the linear increase in time of the mean travelled distance of the quantum walker \cite{ambainisconfproc,Washburn:1992bs}. This ballistic behavior contrasts with the diffusive behavior of a classical particle performing a RW, whose mean travelled distance increases with the square root of time.
Single-particle QWs, as realized in various experiments \cite{MichalKarski07102009,PhysRevLett.104.100503,Schmitz:2009fk,PhysRevLett.100.170506}, can also be mimicked with classical wave mechanics \cite{PhysRevA.69.012310,PhysRevA.68.020301,PhysRevA.61.013410,Gnutzmann:2006fv}.
In contrast, QWs with two particles \cite{PhysRevLett.102.253904, omar:042304, pathak:032351,Peruzzo,0305-4470-39-48-009, zeilingzukow} show purely quantum features due to many-particle interference or entanglement that no classical system can provide. For larger particle numbers, it is unknown how the emerging hierarchy of many-particle interference effects in terms of the number of contributing particles affects physical observables. We will derive a general expression for multi-mode particle number correlation functions which gives access to this hierarchy, and provide first examples as manifest in prominent interference structures in two-mode number correlation functions. While the QW constitutes a paradigmatic testing ground for the study of such many-particle interference, the formalism we present applies to \emph{any} linear, non-interacting scattering scenario.

In a many-particle QW, the distinction between effects of single-particle interference and those due to genuine many-particle interference is crucial. Since distinguishable particles show no many-particle interference, a scenario with non-interacting, distinguishable particles \cite{chandrashekar:022314} is equivalent to many subsequent, independent single-particle walks. Given the position probability distribution of an individual particle, standard combinatorics suffices to derive the many-body counting statistics. Therefore, the differences in the behavior of distinguishable and indistinguishable particles highlight true multi-particle interference. We will see that these differences are not detectable by the mean particle number, but only by observables which capture many-particle correlations. The counting statistics thus encodes an unambiguous signature of many-particle coherence and interference and is readily accessible in the experiment.

\section{Framework}
\begin{figure}[h]
	\includegraphics[width=7cm]{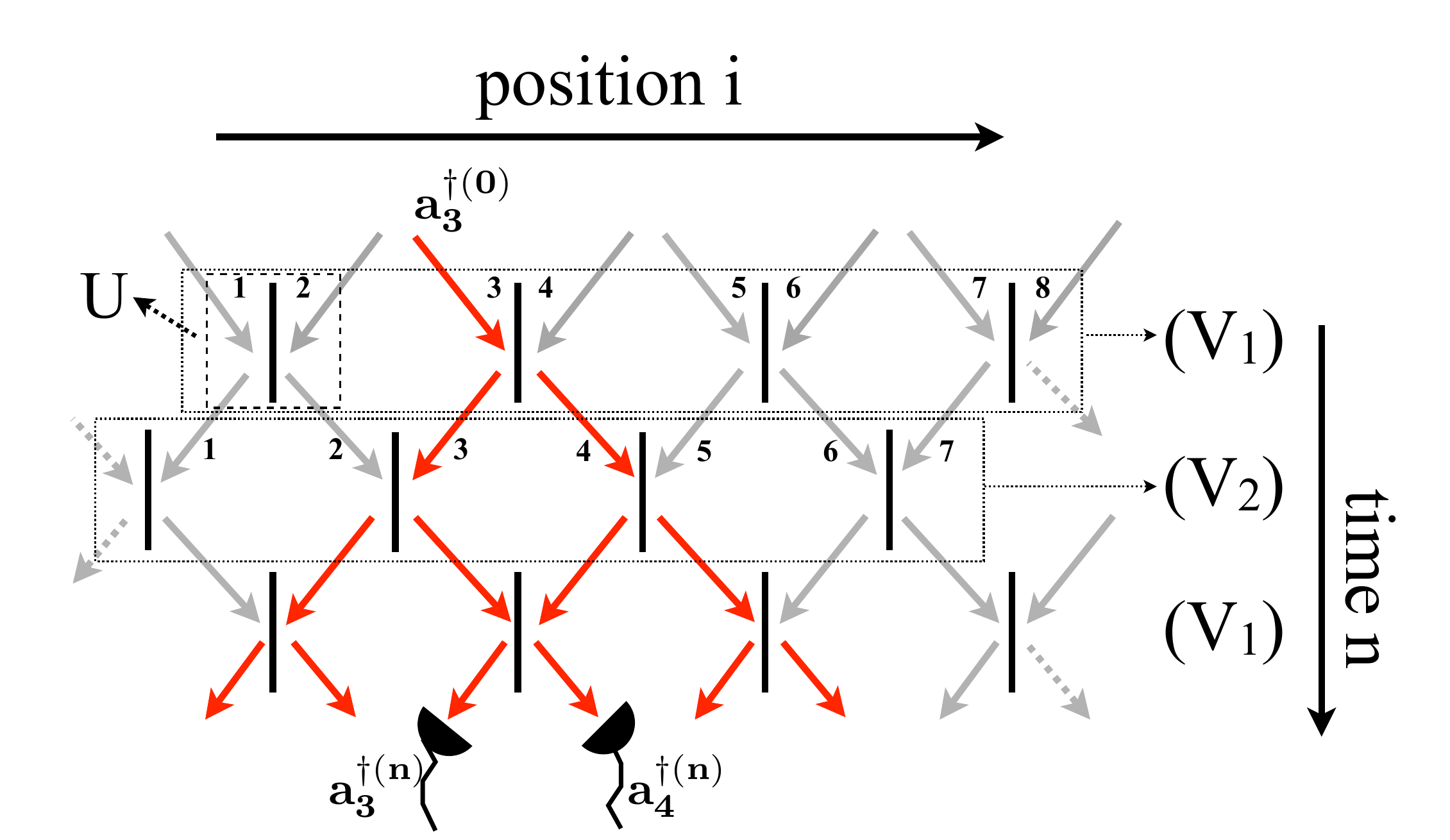}
	\caption{(Color online) Beam splitter array with each BS described by the unitary matrix U, Eq.~(\ref{U}). Particles start in modes with label $(0)$, propagate from top to bottom through the adjacent rows of BS, and are eventually detected in modes with label (n), on exit. }
	\label{arraypic}
\end{figure}

We consider a many-particle QW \cite{rohdemultidtqw} in one dimension, physically realized by N particles which propagate through an array of L unbiased beam splitters (BS) as depicted in Fig.~\ref{arraypic} \cite{1464-4266-1-1-002, PhysRevA.68.032314}. The transverse position of the beam splitters in the array defines the particle position, and subsequent BS rows represent the discrete steps in time.
A particle that passes a single BS will leave it in a coherent superposition of the two output modes, as described by the unitary $2\times 2$ matrix
\begin{equation}
	\mathrm{U}=\frac{1}{\sqrt{2}}\left(\begin{array}{cc}1 & i\\ i & 1\end{array}\right).
	\label{U}
\end{equation}
While the particles propagate through the array, we have to distinguish two inequivalent steps, as depicted in Fig.~\ref{arraypic}: The direct sum $\mathrm{V_1}$ of the single BS matrices U describes odd time steps, where modes 1 and 2 enter one beam splitter, modes 3 and 4 enter the next and so forth. In the following step, modes 2 and 3 couple into the same BS, and so do modes 4 and 5 etc. Even steps are thus described by $\mathrm{V_2=S^-V_1S^+}$, with matrix elements $\mathrm{S^{\pm}_{i,j}}=\delta[\mathrm{(i-j\mp1)_{mod\,2L},0]}$ of $\mathrm{S^{\pm}}$.
The $\mathrm{2L\times 2L}$ matrix
\begin{equation}
	\mathrm{W(n)} = \left\lbrace
		\begin{array}{lc}
			\mathrm{(V_1V_2)^{\frac{n}{2}}}, & \text{for n even}\\
			\mathrm{(V_1V_2)^{\frac{n-1}{2}}V_1}, & \text{for n odd}
		\end{array}\right.
	\label{evop}
\end{equation}
describes the evolution for n time steps.
The creation operator of a particle  $\mathrm{a_j^{\dagger(0)}}$ in mode j at time step 0 is mapped onto a superposition of the creation operators at time step n, with 2L the maximum number of modes:
\begin{equation}
	\mathrm{a_j^{\dagger(0)}} \mapsto \sum\limits_{\mathrm{q=1}}^{\mathrm{2L}}{\mathrm{W_{jq}(n)}\mathrm{a_q^{\dagger(n)}}}.
	\label{unitev}
\end{equation}
Given an N-particle initial state, distributed over 2L input modes,
\begin{equation}
\mathbf{\ket{\psi_{ini}}} = \prod\limits_{\mathrm{j=1}}^{\mathrm{2L}}\frac{1}{\sqrt{\mathrm{r_j!}}}\left(\mathrm{a_j^{\dagger(0)}}\right)^{\mathrm{r_j}}\ket{0},
\label{inistate}
\end{equation}
Eq.~(\ref{unitev}) implies the evolution into the output state
\begin{equation}
	\mathbf{\ket{\psi_{fin}}}=\prod\limits_{\mathrm{j=1}}^{\mathrm{2L}} \frac{1}{\sqrt{\mathrm{r_j!}}} \left(\sum\limits_{\mathrm{q=1}}^{\mathrm{2L}}\mathrm{W_{jq}(n)a_q^{\dagger(n)}}\right)^{\mathrm{r_j}}\ket{0}.
	\label{prodfin}
\end{equation}
In (\ref{inistate}, \ref{prodfin}), $\ket{0}$ and $\mathrm{r_j}$ are the vacuum state and the occupation number of mode j (such that $\rm\sum_{j=1}^{2L}r_j=N$), respectively.

\section{Many-particle observables}
Note that the number of terms in (\ref{prodfin}) grows exponentially with N, so that simulations of the dynamics of many-particle QWs are practically impossible for larger systems. However, expectation values of many experimental observables can be derived without complete knowledge of the full final state. One class of such observables which reveals many-body interference are particle number correlation functions $\mathrm{a_{i_1}^{\dagger(n)}}\dots\mathrm{a_{i_m}^{\dagger(n)}}\mathrm{a_{i_1}^{(n)}}\dots\mathrm{a_{i_m}^{(n)}}$ of m output modes $\mathrm{i_j}$.
For m=1, this is the single-mode particle expectation number, $\mathrm{n_i=a_i^\dagger a_i}$, simply given by the sum of 2L single-particle densities of a single-particle QW,
\begin{equation}
	\langle\mathrm{n_i}\rangle=\sum\limits_{\mathrm{k=1}}^{\mathrm{2L}}{|\mathrm{W_{ki}}|^2\mathrm{r_k}} ~,
	\label{density}
\end{equation}
where $\mathrm{W_{ki}}$ is the transition amplitude from mode k to mode i, and we left out the dependence of $\mathrm{W_{ki}}$ on n, for simplicity of notation. Eq.~(\ref{density}) holds for both, fermions and bosons, as well as for distinguishable particles. Within these different particle-families, various distinct initial states are mapped on final states with the same mean particle number by the beam splitter array. Many-particle interference does not manifest on this level, and no information about the many-particle coherence of the state is obtained, as anticipated above.
This is different when we correlate detection events in two different modes, i.e. m=2 (the upper/lower $\pm$-sign refers to bosons/fermions):
\begin{equation}
	\begin{split}
		&\mathbf{\bra{\psi_{fin}}}\mathrm{a_{i}^{\dagger(n)}}\mathrm{a_{j}^{\dagger(n)}}\mathrm{a_{i}^{(n)}}\mathrm{a_{j}^{(n)}}\mathbf{\ket{\psi_{fin}}}\\& = \sum\limits_{\mathrm{k}<\mathrm{l}=1}^{\mathrm{2L}}{|\mathrm{W_{ki}}\mathrm{W_{lj}}\pm\mathrm{W_{kj}}\mathrm{W_{li}}|^2\mathrm{r_k}\mathrm{r_l}}+\sum\limits_{\mathrm{k}=1}^{\mathrm{2L}}{|\mathrm{W_{ki}}\mathrm{W_{kj}}|^2\mathrm{r_k}(\mathrm{r_k}-1)}.
	\end{split}
	\label{2correlator}
\end{equation}

The two sums in this equation represent two physically distinct processes: in the first, the particles originate from \emph{different} modes. The two-particle amplitude $\mathrm{W_{ki}}\mathrm{W_{lj}}$ describes the simultaneous transition of one particle from mode k to i, and of another one from l to j. Due to the indistinguishability of the particles, this amplitude interferes with that where k and l are exchanged. The second sum accounts for particles that originate from the \emph{same} mode. The corresponding many-particle amplitude $\mathrm{W_{ki}}\mathrm{W_{kj}}$ is invariant upon exchange of the input modes, and hence exhibits no many-particle interference.

Eqs.~(\ref{density}, \ref{2correlator}) exhaust all possible correlations between two particles injected into the BS array. Only higher order correlations between a larger number of particles can exhibit genuine many-particle interference effects.
Since the number of distinct correlations increases like the factorial of the particle number N, a general expression for the m-mode particle number correlation function is desirable, and can be shown to read
\begin{equation}
	\begin{split}
&\langle\mathrm{a_{i_1}^{\dagger(n)}}\mathrm{a_{i_2}^{\dagger(n)}}\dots\mathrm{a_{i_m}^{\dagger(n)}}\mathrm{a_{i_1}^{(n)}}\mathrm{a_{i_2}^{(n)}}\dots\mathrm{a_{i_m}^{(n)}}\rangle_{\mathbf{\ket{\psi_{fin}}}} \\ &= \sum\limits_{\mathrm{q_1}\leq\mathrm{q_2}\leq\dots\leq\mathrm{q_m=1}}^{\mathrm{2L}}  
\prod\limits_{\mathrm{l=1}}^{\mathrm{2L}}{\frac{\mathrm{r_{l}}!}{\mathrm{\tilde{r}_l!}}}
~ \left|~ \sum\limits_{\mu=1}^{\mathrm{S}(\vec{\mathrm{q}})}
{\mathrm{f_{B/F}(\sigma_\mu)}
\prod\limits_{\mathrm{j=1}}^\mathrm{m}\mathrm{W_{{\sigma_{\mu j}},i_j}(n)}
} ~ \right|^2,
	\end{split}
	\label{expval}
\end{equation}
where $\mathrm{\tilde{r}_l} = (\mathrm{r_{l}}-\sum\limits_{\mathrm{k=1}}^{\mathrm{m}}{\delta_{\mathrm{l,q_k}}})$. The function $\mathrm{f_{B/F}}(\sigma_\mu)$ is given by unity for bosons (B), and by the sign of the permutation in its argument for fermions (F).
The outer, incoherent sum in Eq.~(\ref{expval}) runs over all ordered combinations $\vec{\mathrm{q}}=\mathrm{(q_1,..,q_m)}$ of input modes the particles may have originated from. Every single summand represents a physically distinct event. Each such event, in turn, is given by the \emph{coherent} sum over all S($\vec{\mathrm{q}}$) permutations $\sigma_{\mu}$ of $\vec{\mathrm{q}}$, due to the particles' indistinguishability, where $\sigma_{\mu j}$ represents the j-th element of $\sigma_{\mu}$. Clearly, (\ref{expval}) reproduces (\ref{density}) and (\ref{2correlator}) for m=1 and m=2, respectively, with the two sums in (\ref{2correlator}) corresponding to $\mathrm{q_1<q_2}$ and $\mathrm{q_1=q_2}$ in (\ref{expval}).

Expression (\ref{expval}) permits to infer a large variety of many-particle observables, including the particle number counting statistics. 
Since the latter are restricted to occupation numbers zero or one for fermions, we concentrate on bosons in the following.

\section{Counting statistics}
The counting statistics of mode i, i.e. the probability $\mathrm{P^{(i)}(k)}$ to find k particles in mode i, is defined by the linear relation
\begin{equation}
	\mathrm{Q_m^{(i)}}:=\langle\mathrm{a_i^{\dagger(n)}}^{\mathrm{m}}\mathrm{a_i^{(n)}}^{\mathrm{m}}\rangle= \sum\limits_{\mathrm{k}=0}^{\mathrm{N}}{\mathrm{P^{(i)}(k)\frac{k!}{(k-m)!}}}~,\quad 0\leq\mathrm{m}\leq\mathrm{N},
\label{Qone}
\end{equation}
where the $\mathrm{Q_m^{(i)}}$ are determined via Eq.~(\ref{expval}).
This relation can be inverted to yield $\mathrm{P^{(i)}(k)}$, if all moments $\mathrm{Q^{(i)}_m}$ are known. Analogously, the counting statistics $\mathrm{P^{(i,j)}(k_i,k_j)}$ of two modes i and j,
{\it i.e.} the joint probability to find exactly $\mathrm{k_i}$ particles in mode i and $\mathrm{k_j}$ particles in mode j, is given in terms of
\begin{equation}
\begin{array}{ll}
		\mathrm{Q_{m_i,m_j}^{(i,j)}}&\coloneqq\langle\mathrm{a_i^{\dagger(n)}}^{\mathrm{m_i}}\mathrm{a_j^{\dagger(n)}}^{\mathrm{m_j}}\mathrm{a_i^{(n)}}^{\mathrm{m_i}}\mathrm{a_j^{(n)}}^{\mathrm{m_j}}\rangle\\
		&\hspace{3pt}=\sum\limits_{\mathrm{k_i,k_j=0}}^{\mathrm{N}}{\mathrm{P^{(i,j)}(k_i,k_j)\frac{k_i!}{(k_i-m_i)!}\frac{k_j!}{(k_j-m_j)!}}}
		\label{Qtwo}
		\end{array}
\end{equation}
with $0\leq\mathrm{m_i,m_j}\leq\mathrm{N}$. Thus all two-mode moments $\mathrm{Q^{(i,j)}_{m_i,m_j}}$ are required for the two-mode counting statistics. Multi-mode counting statistics for more than two modes are obtained in strictly analogous fashion.

\subsection{Single-mode statistics}
\begin{figure}[h]
	\includegraphics[width=7.5cm]{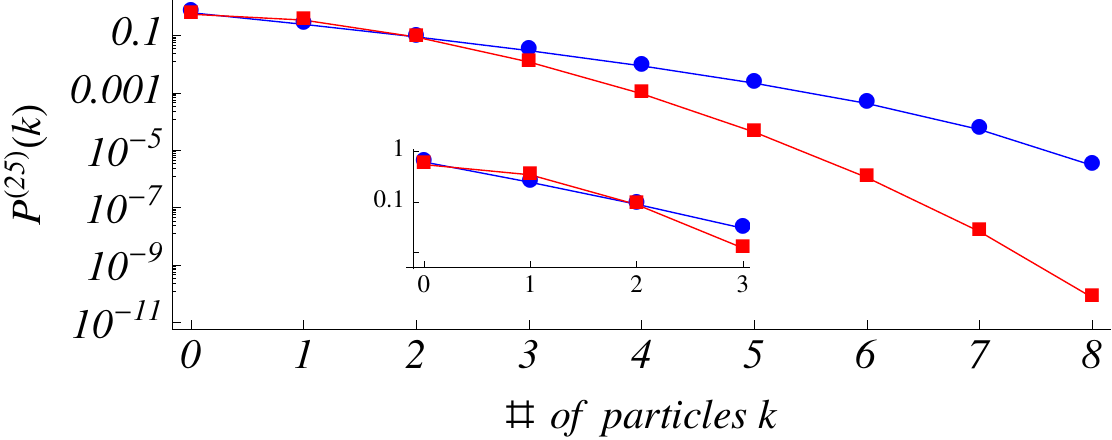}
	\caption{(Color online) Single-mode counting statistics P$^{(25)}$(k) for mode i=25 after $\mathrm{n=6}$ steps on a logarithmic scale. Bosonic bunching enhances P$^{(25)}$(k) for bosons by several orders of magnitude (blue circles) as compared to the result for distinguishable particles (red squares). The inset shows the distribution for small k.}
	\label{onesite}
\end{figure}

With these tools, we can now fully characterize the distribution of particles among the modes, and work out the observable consequences of many-particle interference: We start out with N=8 particles, launched in eight adjacent modes at the center of an array of length 2L=50, and initially in a product state (\ref{inistate}). 
Fig.~\ref{onesite} shows the probability for mode 25 to be occupied by k  particles after $\mathrm{n=6}$ steps.
The probability decreases with k, both for bosons and distinguishable particles. Since, however, for bosons all many-particle amplitudes that contribute to the coherent sum in (\ref{expval}) are identical, they interfere constructively. Consequently, the probability to find more than two particles in one mode decreases much slower for bosons than for distinguishable particles. For the latter, P$^{(25)}$(k) exhibits Poissonian decrease like 1/k!. In contrast, in the limit of many particles, P$^{(25)}$(k) can be shown to approach exponential decrease in the bosonic case, which is apparent in Fig.~\ref{onesite} for low and intermediate values of k. For k=N=8, P$^{(25)}$(k) is equal to $\rm Q^{(25)}_{k=N=8}$ in (\ref{Qone}). There, all N! amplitudes in the coherent sum in (\ref{expval}) are necessarily equal. Thus, bosonic bunching \cite{PhysRevLett.59.2044,PhysRevLett.104.220405} induces a factor N!=8! as compared to distinguishable particles. For $\rm k<N$, P$^{(25)}$(k) is a sum of several $\rm Q_m^{(25)}$, what prevents such simple derivation of the bosonic value of P$^{(25)}$(k).

Further note that, since the number of constructively interfering amplitudes grows monotonically with the number of detected particles, the distribution in Fig.~{\ref{onesite}} is smooth (in contrast to the strong fluctuations in Fig.~\ref{twomodecut} below), and, in particular, it is qualitatively independent of the considered output mode. 

\subsection{Two-mode statistics}
\begin{figure}
	\includegraphics[width=8cm]{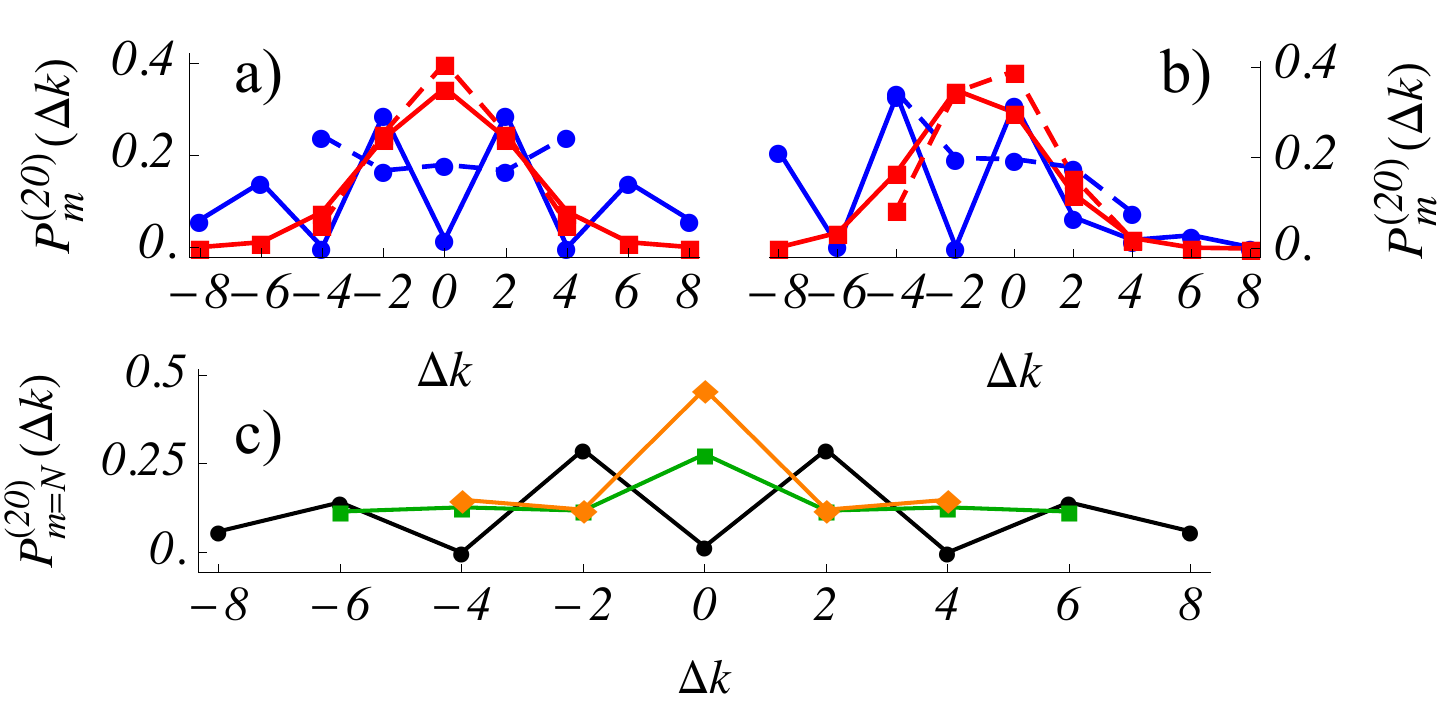}
	\caption{(Color online) Joint counting statistics for two different pairs of modes: Particle distribution $\mathrm{P_{m}^{(20)}(\Delta\rm k =k_i-k_j)}$ among two modes after n=20 steps of N=8 particles, for lattice size 2L=50, conditioned on finding m=4 or 8 particles. a) Joint probability distribution of the symmetric (with respect to the lattice center) modes i=19, j=32 to find m=4 particles (dashed; distinguishable: red squares, bosons: blue circles) and m=8 particles (solid; distinguishable: red squares, bosons: blue circles). b) Same as a), for asymmetric modes i=18, j=32. c) $\mathrm{P_{m=N}^{(20)}(\Delta\rm k =k_i-k_j)}$ of modes i=19, j=32 for N=4 (orange diamonds), N=6 (green squares) and N=8 (black circles).}
	\label{twomodecut}
\end{figure}

In contrast to the above single-mode counting statistics, \emph{distinct} many-particle amplitudes contribute to the coherent sum in (\ref{expval}) when \emph{several} output modes are correlated, and give rise to characteristic many-particle interference effects beyond the smooth bunching observed in Fig.~\ref{onesite}. For these interference effects to fully develop, we need to propagate the many-particle state over a larger number of steps (n=20), such that all single particle wave functions spread over the entire lattice, and mode-to-mode correlations can build up. Our figure of merit is then the conditional probability $\mathrm{P_m^{(n)}(k_i,k_j)}$ to find $\mathrm{k_{i/j}}$ particles in mode $\rm i/j$, with a total of $\rm m=k_i+k_j$ particles in these modes. Indeed, as evident in Fig.~\ref{twomodecut}, many-particle interference of indistinguishable particles has a marked impact on $\mathrm{P_{m=N=8}^{(20)}(k_i,k_j)}$ as compared to the case of distinguishable ones.
While the latter always implies a single-peaked dependence of $\mathrm{P_{m=N=8}^{(20)}(k_i,k_j)}$ on the particle imbalance $\Delta\rm k=k_i-k_j$, many-particle interference manifests in strong modulations of $\mathrm{P_{m=N=8}^{(20)}(k_i,k_j)}$ for bosons, with pronounced maxima for even values of $\Delta\rm k$, i.e. for odd particle numbers in either one of both modes (remember that N=8). Furthermore, the many-particle interference condition is strongly affected by the specific choice of the output modes i and j, as obvious from a comparison of Figs.~\ref{twomodecut}a) and b): an asymmetric choice of i and j with respect to the lattice center induces a strong asymmetry in $\mathrm{P_{m}^{(20)}(k_i,k_j)}$.
What's more, the interference signal sensitively depends on the total number of particles N in the system, in a non-monotonic way, as apparent from Fig.~\ref{twomodecut}~c), where the bosonic signal is shown for different N. This allows for a clear distinction of multi-particle interference contributions of different order.

When not all particles contribute to the signal in modes i and j, i.e. for $\rm m=k_i+k_j<N$, many of the events which contribute to $\mathrm{P_{m<N}^{(20)}(k_i,k_j)}$ can be distinguished by the final positions of the remaining $\rm(N-m)$ particles and thus cannot interfere anymore, what leads to a smoother dependence of $\mathrm{P_{m<N}^{(20)}(k_i,k_j)}$ on $\rm\Delta k$. This is clearly apparent in Figs.~\ref{twomodecut}~a),b), as well as from (\ref{expval}): for expectation values that involve the maximum number $\mathrm{N=8}$ of creation/annihilation operators, the incoherent sum collapses to a single term while the coherent sum contains up to N! terms. For $\mathrm{m<N}$, many more incoherently added terms contribute, and, thus, interference effects are averaged out.

\begin{figure}
	\includegraphics[width=8cm]{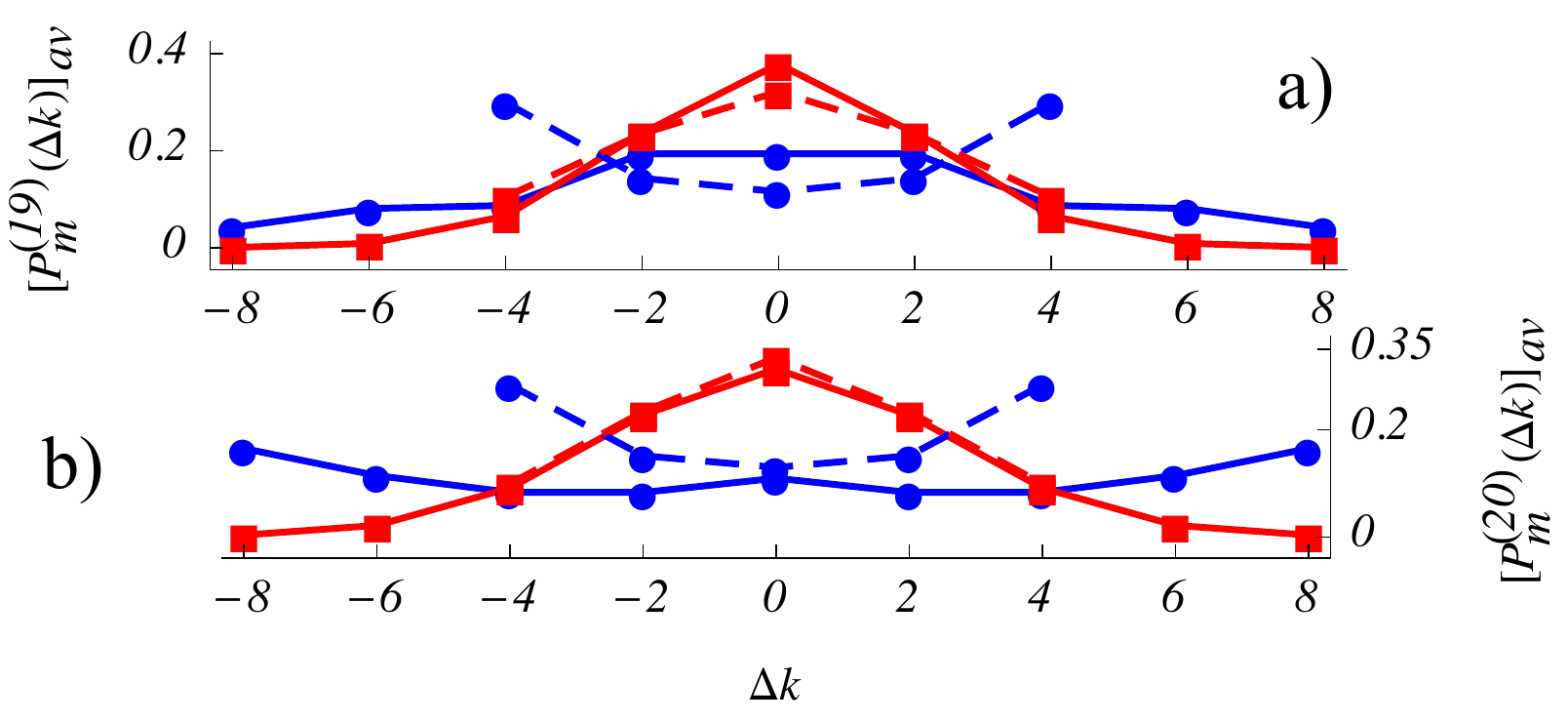}
	\caption{(Color online) Two-mode statistics as in Fig.~\ref{twomodecut}, averaged over all pairs of modes, after a) n=19 steps and  b)  n=20 steps. Color code as in Fig.~\ref{twomodecut}. Note that while the distribution is essentially invariant under changes of n for distinguishable particles, the correlations fluctuate with n, in the bosonic case.}
	\label{twomodecutmean}
\end{figure}

Let us finally stress that the observed many-particle interference effects even persist after an average of $\mathrm{P_{m=N=8}^{(20)}(k_i,k_j)}$ over all pairs $\rm(i,j)$: Fig.~\ref{twomodecutmean} shows that the bosonic signal is markedly different from the classical result, after a given number of steps as well as in its dependence on n. While the two-mode correlation is single-peaked and essentially n-\emph{independent} for distinguishable particles, some of the interference structure of Fig.~\ref{twomodecut} prevails under the average, and, furthermore, shows considerable variation with n. This latter dependence is, similarly to the case considered in Fig.~\ref{twomodecut}, reduced for $\rm m<N$.

\section{Conclusion}
In summary, many-particle interferences between N indistinguishable particles injected into the 2L input modes of a beam splitter array give rise to strong bunching effects in the single mode counting statistics, as well as to a robust, pronounced and non-monotonic dependence of the two-mode correlation function on total particle number and particle imbalance between both modes. The many-particle interference correction induces large deviations from the classical (distinguishable particle) result. This allows to characterise the witnessed interference phenomenon in terms of the number of contributing particles -  an information unaccessible, e.g., in few-particle quantum walks. This is of particular interest, e.g., in the many-particle transport across disordered media \cite{YoavLahini}, which is also covered by our central result, Eq.~(\ref{expval}), that actually allows to explore the full abundance of statistical observables of any non-interacting system with an arbitrary number of particles.

The authors acknowledge fruitful and stimulating discussions with P. M. Dibwe and F. Petruccione. Partial financial support by the German Federal Ministry for Science
and Education under the grant SUA 08/008 is gratefully acknowledged. M.C.T. acknowledges financial support by Studienstiftung des deutschen Volkes.


\begin{thebibliography}{10}

\bibitem{PhysRevA.48.1687}
Y.~Aharonov, L.~Davidovich, and N.~Zagury,
\newblock Phys. Rev. A {\bf 48}, 1687 (1993).

\bibitem{Kempe:2003fk}
J.~Kempe,
\newblock Contemp. Phys. {\bf 44}, 307 (2003).

\bibitem{ambainisconfproc}
A.~Ambainis, {\it et al.},
\newblock Proc. STOC'01, Heraklion (Greece), 37 (2001).

\bibitem{Washburn:1992bs}
S.~Washburn and R.~A. Webb,
\newblock Rep. Prog. Phys. {\bf 55}, 1311 (1992).

\bibitem{MichalKarski07102009}
M.~Karski, {\it et al.},
\newblock Science {\bf 325}, 174 (2009).

\bibitem{PhysRevLett.104.100503}
F.~Z\"ahringer, {\it et al.},
\newblock Phys. Rev. Lett. {\bf 104}, 100503 (2010).

\bibitem{Schmitz:2009fk}
H.~Schmitz, {\it et al.},
\newblock Phys. Rev. Lett. {\bf 103}, 090504 (2009).

\bibitem{PhysRevLett.100.170506}
H.~B. Perets, {\it et al.},
\newblock Phys. Rev. Lett. {\bf 100}, 170506 (2008).

\bibitem{PhysRevA.69.012310}
H.~Jeong, M.~Paternostro, and M.~S. Kim,
\newblock Phys. Rev. A {\bf 69}, 012310 (2004).

\bibitem{PhysRevA.68.020301}
P.~L. Knight, E.~Rold\'an, and J.~E. Sipe,
\newblock Phys. Rev. A {\bf 68}, 020301 (2003).

\bibitem{PhysRevA.61.013410}
D.~Bouwmeester, {\it et al.},
\newblock Phys. Rev. A {\bf 61}, 013410 (1999).

\bibitem{Gnutzmann:2006fv}
S.~Gnutzmann and U.~Smilansky,
\newblock Adv. Phys. {\bf 55}, 527  (2006).

\bibitem{PhysRevLett.102.253904}
Y.~Bromberg, {\it et al.},
\newblock Phys. Rev. Lett. {\bf 102}, 253904 (2009).

\bibitem{omar:042304}
Y.~Omar, {\it et al.},
\newblock Phys. Rev. A {\bf 74}, 042304 (2006).

\bibitem{pathak:032351}
P.~K. Pathak and G.~S. Agarwal,
\newblock Phys. Rev. A {\bf 75}, 032351 (2007).

\bibitem{Peruzzo}
A.~Peruzzo, {\it et al.},
\newblock Science {\bf 329}, 1500 (2010).

\bibitem{0305-4470-39-48-009}
M.~Stefanak, {\it et al.},
\newblock J. Phys. A {\bf 39}, 14965 (2006).

\bibitem{zeilingzukow}
K.~Mattle, {\it et al.},
\newblock Appl. Phys. B {\bf 60}, S111 (1995).

\bibitem{chandrashekar:022314}
C.~M. Chandrashekar and R.~Laflamme,
\newblock Phys. Rev. A {\bf 78}, 022314 (2008).

\bibitem{rohdemultidtqw}
P.~P. Rohde, {\it et al.},
\newblock arXiv:1006.5556, (2010).

\bibitem{1464-4266-1-1-002}
P.~T{\"o}rm{\"a} and I.~Jex,
\newblock J. Opt. B {\bf 1}, 8 (1999).

\bibitem{PhysRevA.68.032314}
M.~Hillery, J.~Bergou, and E.~Feldman,
\newblock Phys. Rev. A {\bf 68}, 032314 (2003).

\bibitem{PhysRevLett.59.2044}
C.~K. Hong, Z.~Y. Ou, and L.~Mandel,
\newblock Phys. Rev. Lett. {\bf 59}, 2044 (1987).

\bibitem{PhysRevLett.104.220405}
M.~C. Tichy, {\it et al.},
\newblock Phys. Rev. Lett. {\bf 104}, 220405 (2010).

\bibitem{YoavLahini}
Y.~Lahini, {\it et al.},
\newblock Phys. Rev. Lett. {\bf 105}, 163905 (2010).

\end{thebibliography}
\end{document}